# Hub Identification of the Metro Manila Road Network Using PageRank

Jacob CHAN[1], Kardi TEKNOMO[2]

[1]Department of Information Systems and Computer Science, School of Science and Engineering
Ateneo de Manila University, Loyola Heights, Quezon City, Philippines 1108
Telephone +632-426-6001, Fax. +632-4261214
E-mail: jtchan@ateneo.edu

[2]Department of Information Systems and Computer Science, School of Science and Engineering
Ateneo de Manila University, Loyola Heights, Quezon City, Philippines 1108
Telephone +632-426-6001, Fax. +632-4261214
E-mail: teknomo@gmail.com

**Abstract**

We attempt to identify the different node hubs of a road network using PageRank for preparation for possible random terrorist attacks. The robustness of a road network against such attack is crucial to be studied because it may cripple its connectivity by simply shutting down these hubs. We show the important hubs in a road network based on network structure and propose a model for robustness analysis. By identifying important hubs in a road network, possible preparation schemes may be done earlier to mitigate random terrorist attacks, including defense reinforcement and transportation security. A case study of the Metro Manila road network is also presented. The case study shows that the most important hubs in the Metro Manila road network are near airports, piers, major highways and expressways.

## 1. Introduction

Roads are important access points because connects different places like cities, districts, and landmarks. Studying road networks is important because disrupting at least one road can affect the connectivity of a road network, the safety of the people, and the accessibility of each road [1].

Studying the robustness of a network is a relatively new research area. Generally, network robustness is defined as the ability of a system to operate correctly under a wide range of change in operational conditions, and to fail gracefully under these conditions by tolerating sudden changes caused by different scenarios [6,10]. However, in transportation studies, road network robustness is the ability of the system to keep a certain capacity level to handle traffic situations under abnormal conditions. Table 1 shows a wide-ranging overview of several road network robustness indices and measures and is compared to one another based on their approach, strengths, and weaknesses.

**Table 1** Comparative analysis of different methodologies on network robustness indices

| Author | Method | Strength | Weakness |
|---|---|---|---|
| Scott, et.al [14] | Network Robustness Index | Better than V/C model | Cannot handle isolated nodes |
| Wakabayashi and Fang [20] | Cost-Benefit Analysis with different indices | Useful reliability index | Useful for limited networks only |
| Tampère, et.al [16] | Dynamic Traffic Assignment | Realistic | Incomplete methodology |
| Knoop, et.al [8] | Incidence Impact | Good correlation | Not good for freeways |
| de Olivera, et.al [3] | Vulnerability and Congestion Indicators | Good vulnerability indicators | Computation time |
| Murray, et.al [12] | Multi-methodology | Scenario identification | Biased |

However, the indices used in analyzing road robustness did not consider random attacks on the network, specifically man-made disasters like terrorist attacks. Terrorist attacks pose different issues not found in natural disasters because they are difficult to protect for law enforcement agencies [19]. Furthermore, aggressors are motivated, and thus exploit schedules in order to get around patrol schedules from law enforcement agencies. As a result, these agencies randomize their scheduling techniques. However, these agencies need to protect thousands of vehicles from terrorist attacks, which pose different risks like threat, vulnerability, and consequence [21].

In this paper, we propose a model that uses PageRank to identify the most important hubs in a road network, where each hub has a certain level of importance. Knowing these hubs may contribute to awareness of safety during terrorist attacks. In Section 2, we discuss the important terms used for the study, as well as different literature on road robustness analysis. Section 3 deals with how our model was designed. We present a case study in the Metro Manila road network in Section 4. Finally, in Section 5, we discuss the conclusions we have from our case study, as well as possible future works regarding road network analysis.

## 2. Literature Review

In this section, we define the terminologies used for our research to better understand our proposed model. We also present research on terrorist attacks and its effects on transportation networks, as well as the concept of PageRank and its variants.

### 2.1. Road Networks

A road network is a weighted directed graph where links represent actual roads, and nodes represent intersections and corners. The weight of each link represents the distance of the link from one source node to a destination node. Each link can only be traversed in one direction only, depending on where the destination node the link is directed. Road networks differ from another in terms of network topology and link distances. Road networks can be represented in several ways, including images from geographical information systems and adjacency matrices.

### 2.2. Network Connectivity

In graph theory, a strongly connected network is a property where each node is reachable from any node, regardless of weight or distance. Tarjan [17] proposed an algorithm that finds the strongly connected components in a graph using depth-first search. A graph is divided into sub-graphs, where each component in each sub-network is strongly connected. Tarjan's algorithm has a runtime of O (|V| + |E|), where |V| represents the number of nodes in the network, while |E| represents the number of links. However, the runtime of the algorithm was reduced to O(|V|) using a binary decision technique [5].

### 2.3. Terrorist Attacks and Robustness

Natural disasters like earthquakes and landslides have been used to find possible evacuation routes and assess road networks before such disasters happen [7,13]. However, planned disasters like terrorist attacks pose different issues because unlike natural disasters, they are less uniform due to their unpredictability [9]. Furthermore, transportation networks and public transits are the primary target of most terrorist attacks because it an easy and accessible way to take lives, cause damage, spread fear, and impact local, regional, and national economies [4].

Willis, et.al [21] estimated the risks involved in terrorist attacks through quantification. He used threat, vulnerability, and consequences, and multiplied it to get an estimation of terrorism risks, which are probabilistic. Threat is the probability that a target will be attacked at a certain time in a certain way. Vulnerability is probability that an attack results in damages, given that an attack occurs. Lastly, consequence is the magnitude of damage given a specific attack type.

### 2.4. PageRank

PageRank is a technique used to determine the importance of nodes in a network, and is applied in search engines like Google [2]. A website node in a web network is considered important when referenced by other websites. The more that the website is cited, the more likely that it is important. In short, PageRank measures the inflow of the nodes and ranks each node based on the number of inflowing links.

For transportation networks, a method called NodeRank [11] was created based on the PageRank algorithm. Three standards were used to measure the degree of importance of each node, namely the topology, weights of each link, and importance of other vertices it connects to. Part of traffic flow information can be acquired using this technique. Using this technique can also assist in traffic system analysis, safety evaluation, and optimization.

## 3. Methodology

We divided our network structure into two major parts. The first part is cleaning the data with Tarjan's algorithm. We then identified the most significant nodes called hubs using PageRank. Finally, we used our model in a case study of the Metro Manila road network, and studied and analyzed our results.

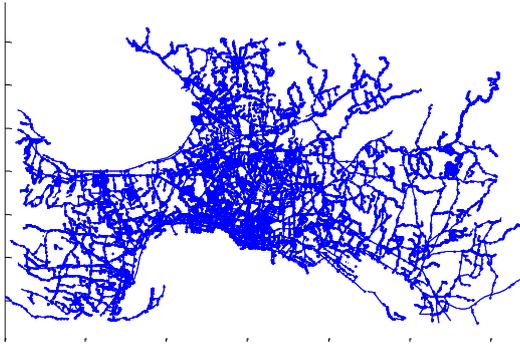

**Fig. 1** Metro Manila Road Network

### 3.1. Distance Calculation

We collected a road network data and mapped it as nodes and links using a distance matrix **D**, which has n rows and n columns, where n respresents the number of nodes. Each node in the data set has a node ID and a longitude-latitude value, and represents an intersection in the entire road network. The nodes are arranged by node ID, and are mapped as an index in **D**.

Each link connects exactly two nodes in the network, and are placed in a designated row and column address to represent its corresponding source and destination node, respectively. The weight of each link was calculated using the Haversine [15] distance formula because it remains well conditioned even at small distances. The Haversine formula can be expressed as a Haversine function, as shown in the following equation:

$$\text{hav}(\theta) = \sin^2\left(\frac{\theta}{2}\right) \quad (1)$$

Given two points with a corresponding latitude and longitude coordinates in degrees, denoted by ($\phi_1$, $\lambda_1$) and ($\phi_2$, $\lambda_2$), respectively, we first converted the coordinate values to radians. Then we computed the distance $d$ (in kilometers) of each link, and stored it in row $i$ and column $j$ of matrix **D**, denoting the source and destination nodes as shown in this equation:

$$d_{ij} = 2r \sin^{-1}\left(\sqrt{\sin^2\left(\frac{\Delta\phi}{2}\right) + \cos\phi_1 \cos\phi_2 \sin^2\left(\frac{\Delta\lambda}{2}\right)}\right) \quad (2)$$

The value of $\Delta\phi$ and $\Delta\lambda$ represent the change in latitude and longitude values of the node pair, respectively. The resulting distance is put in row $i$ and column $j$ of matrix **D** to indicate a link between node $i$ and node $j$. However, the value of the $d_{ij}$ is 0 if no link exists between node $i$ and node $j$.

### 3.2. Data Cleaning

Once the distance values have been computed for each element in matrix **D**, we cleaned our data with Tarjan's algorithm [17]. The network was divided into smaller groups known as sub-networks that represented the strongly connected nodes in the network. The sub-network with the most number of strongly connected components was retained in the network, while the sub-networks with less number of strongly connected nodes were considered weaker sub-networks, and were immediately removed from the network. We considered the strongly connected nodes and links because they would greatly affect the connectivity and topology of the entire network. The largest strongly connected sub-network was used for analysis of our case study, as shown in Figure 1. The new network is represented as a distance matrix **D**$_{new}$, which is an m by m matrix, where m represents the number of nodes in the strongly connected network.

### 3.3. PageRank

After cleaning the data, we performed PageRank on the network through matrix **D**$_{new}$. Given the matrix **B**, which represents the adjacency matrix version of **D**$_{new}$, we computed the probability matrix **P**, which is the probability of each element over the sum of the outbound links, represented by the element $j$. Matrix **P** has the same dimensions as matrix **B** and **D**$_{new}$. The following equation further explains this:

$$p_{ij} = \frac{b_{ij}}{\sum_{i=1}^{m} b_{ij}} \quad (3)$$

The value of $m$ is the number of nodes in the network, which are the dimensions of matrix **B**. Next, we computed the sum-product $sp$ of each row using the following formula:

$$sp_i = \sum_{j=1}^{m} p_{ij} * c_j \quad (4)$$

The value of $c_j$ represents the number of outbound nodes in the network. After computing for the sum-product value of each row, we calculated the value of PageRank $c_i$ for each row for the next iteration using the recursive rank formula:

$$c_i = (1-d) + d*sp_i \quad (5)$$

The variable $d$ is an adjustable parameter. We chose the value of 0.85 because it was the default value in Teknomo's [18] PageRank algorithm. The PageRank algorithm will run until convergence, where the sum of $c_i$ is equal to the number of nodes in the network. The results of PageRank always have the same ranks for a number of iterations. This is known as recursive ranking, which is an optimization technique. Teknomo [18] gives a simulation and more details on PageRank for further understanding on how the technique works.

Once the values of $c$ have been computed, their normalized values were computed using the following formula:

$$norm_i = \frac{c_i}{\sum_{i=1}^{m} c_i} \quad (6)$$

When each normalized value was computed, the nodes were sorted and ranked according to their normalized values. A standard competition ranking was used for ranking the normalized values. The tied values were assigned exactly the same rank values, and then the next element after the tied elements was left with a gap in the ranking. A higher normalized value means that the node has more impact to the network, thus a higher rank is assigned. However, a lower normalized value means that the node was less important.

### 3.4 Hub Location

Once we gathered the necessary data, we proceeded with hub identification and location. Using the longitude and latitude data we gathered from OpenStreet Map (OSM), we located the most important hubs using Google Maps. We also counted the in-degree and out-degree nodes, and their total values to be compared in line with PageRank. We multiplied the normalized values by 1,000,000 to measure the *strength* of the node, and correlated it with the rank in the network. We also did further analysis in the Metro Manila network.

### 4. Results and Analysis

The Metro Manila road network was used for the case study of our model. The data was collected using OSM. A total of 69,577 nodes were found in the Metro Manila network. When Tarjan's algorithm was run to clean the network, the number of nodes was reduced to 66,854. PageRank was used to rank the importance of each node. The PageRank algorithm of Teknomo [18] was run for 204 iterations to produce the ranks of each node in the road network.

### 4.1. Identifying the Highest Ranked Nodes

We collected the 25 most important node hubs in the Metro Manila road network by plotting their respective latitude-longitude values to Google Maps in order to find their exact location, and which area is approximately the nearest among those hubs. Table 2 shows the 25 most important hubs based on PageRank. The landmarks were included because we wanted to know which areas does each location capture. We did not assume anything in determining the landmarks, except the local knowledge we had about these places.

**Table 2** First 25 Ranked Nodes According to PageRank

| Node Strength | Location | Landmark/s |
|---|---|---|
| 69.96 | Claveria, Binondo, Manila | Ongpin China town |
| 65.70 | Tomas Pinpin, Binondo, Manila | Ongpin China town |
| 62.88 | P.Burgos Ave., Ermita, Manila | Manila City Hall/Philippine Normal University |
| 57.82 | Valenzuela, Quezon City | North Luzon Expressway |
| 57.65 | P.Burgos Ave., Ermita, Manila | Manila City Hall/Philippine Normal University |
| 55.634 | Sta. Maria, Bulacan | Sacred Heart Academy/Sta. Maria Elem.School |
| 52.77 | EDSA-Greenhills, San Juan | Corinthian Gardens |
| 52.31 | Newport, Pasay | Manila International Airport 3/Villamor Air force base/Resorts World/Marriott Hotel |
| 51.33 | Dasmariñas, Binondo, Manila | Ongpin China town |
| 48.47 | Valenzuela, Quezon City | North Luzon Expressway |
| 48.07 | C5-Taguig | West Bicutan National HS/C5 Highway/Heritage Park |
| 46.30 | Bacoor, Cavite | Coastal Road/Bali Garden/Manila-Cavite Expressway |
| 46.17 | Santa Cruz, Manila | Carriedo Train Station |
| 45.82 | Ermita, Manila | Paco Park/Manila Science HS |
| 45.49 | Guiguinto, Bulacan | McArthur Highway |
| 45.41 | Sta. Teresita, Quezon City | SM Sun Residences/Welcome Roundabout |
| 44.65 | EDSA Highway-Ortigas Flyover | Corinthian Gardens |
| 44.38 | Greenhills, San Juan | EDSA Highway |
| 43.97 | Bacoor, Cavite | Coastal Road/Bali Garden/Manila-Cavite Expressway |
| 43.8104675 | Balintawak, Quezon City | Cry of Balintawak/North Luzon Expressway/EDSA Highway |
| 43.42223344 | Ayala Avenue, Makati | Glorietta 5/InterContinental Manila/SM Makati/EDSA Highway |
| 43.36666459 | Sampaloc, Manila | Legarda Elementary School/Aldana School |
| 43.30983927 | Barangka, Marikina | Riverbanks/Barangka Elementary School |
| 43.26685007 | San Bartolome, Quezon City | Quirino Ave/Mindanao Ave/Quirino Roundabout |
| 42.76296099 | Marcelo Green, Parañaque | Osmeña Highway/Skyway/Pasay Train Station |

Out of the 25 highest ranked hubs, 8 were located in Manila. Out of the 10 highest ranked hubs, 5 were located in Manila. Based on these findings, we found out that Metro Manila has some of the most important roads in the network. Cutting these roads may impact the connectivity of the entire network. We also found out that the landmarks from the 25 most important hubs were mostly found public places like schools, malls, business districts, or even at major roads, especially the North Luzon Expressway (NLEX), EDSA, and the C5 Roads, as shown in Figures 2 and 3.

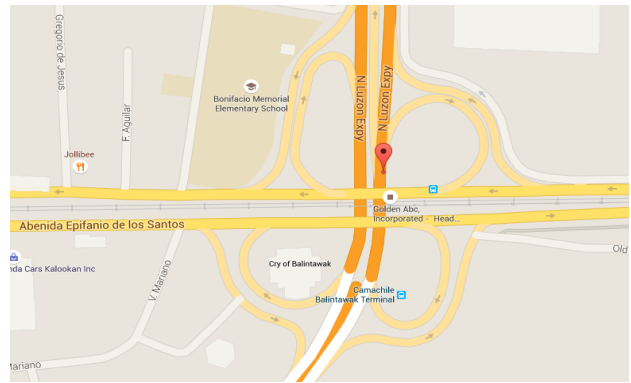

**Fig. 2** EDSA-Balintawak-NLEX Interchange

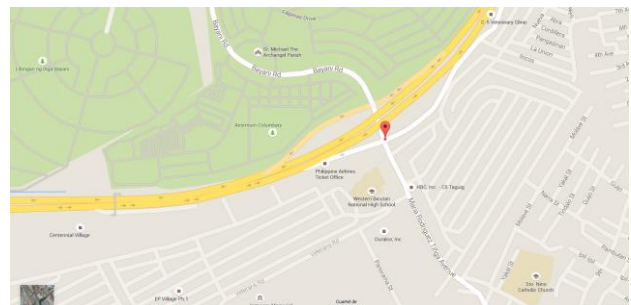

**Fig. 3** C5-Taguig Highway Near Heritage Park

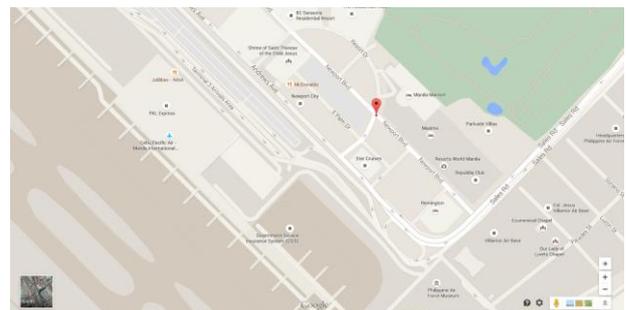

**Fig. 4** Manila International Airport Terminal 3, Pasay City

Figure 4 shows a hub near Terminal 3 of the Manila International Airport, which is near the Philippine Air Force military base. PageRank was able to identify this as a hub, despite that only topology was used to find out the important hubs.

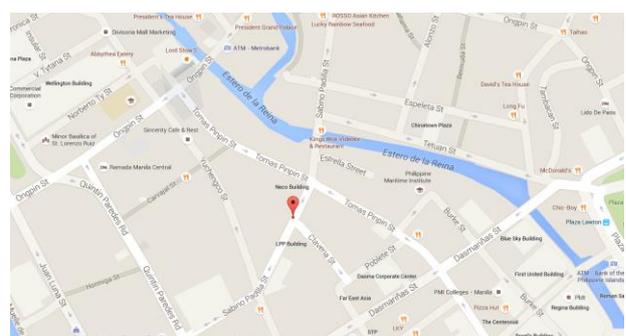

**Fig. 5** Binondo Chinatown Hub, Manila

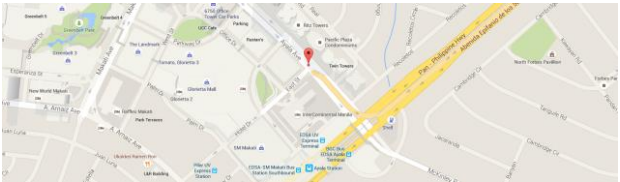

**Fig. 6** Ayala Ave. Business District Hub, Makati

Figure 5 shows a hub in Binondo, Manila. This is a place where trade is most prominent since the place is known as the Chinatown of Manila. The place is also near a flea market and a mall. According to PageRank, this was the most important hub in the Metro Manila road network. Again, PageRank was able to identify the important hubs given the topology alone.

Figure 6 shows a hub in Makati, which is the central business district in Metro Manila. This is also important to note because incidentally, a bombing occurred in one of the five Glorietta malls last 2007. In other words, the Makati road network is very important to mitigate sooner based on PagRank. However, this was not the case all the time, as shown in Table 3.

**Table 3** Bottom 5 Lowest Ranked Hubs

| Strength | Location | Landmark |
|---|---|---|
| 3.55 | Kalayaan Ave, Makati | Bel-Air |
| 3.55 | Makati Ave., Makati | Bel-Air |
| 3.55 | Kalayaan Ave, Makati | Bel-Air |
| 3.63 | Meycauayan, Bulacan | McArthur Highway |
| 3.63 | Meycauayan, Bulacan | McArthur Highway |

The lowest ranked hubs were also located in Makati, specifically Kalayaan Avenue, which is one of the busiest streets in the city. Even though Makati is a business district, the least ranked hubs were located here perhaps because most roads in Makati are one-way. All the locations specified by Table 3 were one way. The two roads in the province of Bulacan were also one-way, which means that these roads were not a threat to be attacked by terrorists according to PageRank. Makati would more likely be attacked because it is a business district, but topologically, in the Bel-Air area, it has the lower rank, as shown in Figure 7.

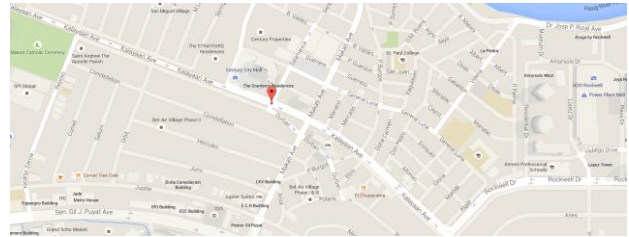

**Fig. 7** Makati-Bel Air Hub

There were also important hubs located outskirt of Metro Manila, like there was one in Bulacan and one in Cavite, which are not included as part of the National Capital Region. Perhaps, this was also a result of the data gathered by OSM, or the processing during data cleaning with Tarjan's algorithm.

### 4.2. Hub Analysis

The strength of each normalized value was computed in order to assess its impact to the road network on a larger scale. The results of the strength were plotted to the rank of each node, as shown in Figure 8.

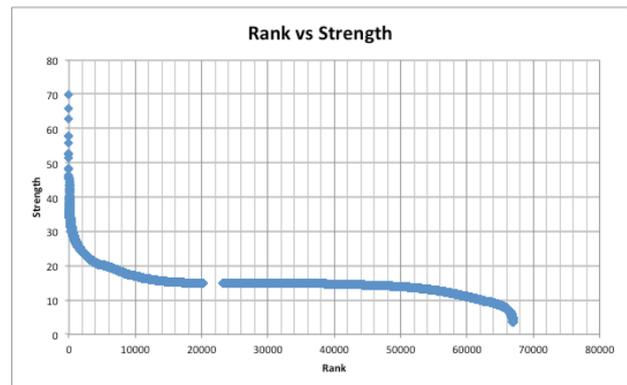

**Fig. 8** Comparing Rank and Strength of Each Node

A sigmoid-like function can be seen in Figure 8. The highest strength value was 69.97, which is located at Binondo, Manila. The lowest strength value was 3.55, which was the strength value located at Kalayaan Avenue, Makati. Holes were present in some parts of the graph, which represents the ties in the ranking. The normalized values and their ranks were logarithmic in nature. The first few ranks were nearer to one another, yet had different strength values and decreased gradually, signifying that these nodes were very important nodes compared to the rest of the network. As the ranks increased, the strength values began to have fewer changes, implying that the discrepancies of the strength values among nodes became smaller, even though a downward movement can still be seen. The values began to subside at the strength value of approximately 15. In other words, the important nodes were highly clustered together, which signifies the importance of each node in the network.

Figure 9 illustrates the trend of the first 25 nodes shown in Table 2. The first 11 hubs had big discrepancies in terms of their strengths, and as the rank increases, the strength starts to converge linearly, meaning that nodes began to be closer to one another. This was evident starting from Node 12 up to 25. In relation to Figure 8, the trends do not necessarily have to dramatically drop, but they get reduced little by little until the lowest rank is achieved. In other words, Figure 9 supports the claim we made in Figure 8.

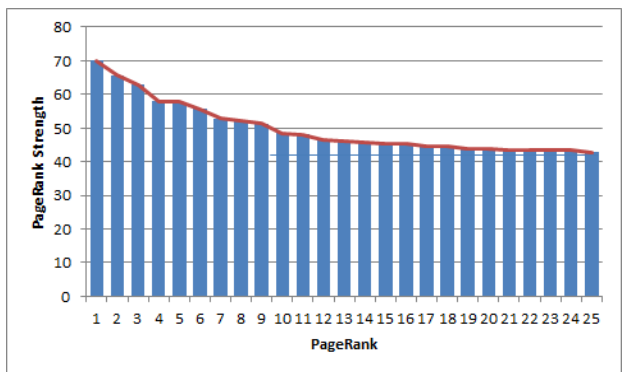

**Fig. 9** PageRank Strength and Actual PageRank Comparison

Another interesting trend is found in Figure 10. Both PageRank and PageRank strength explained the outflowing nodes with 35% and 39% R-square values, respectively for linear and exponential model. This implies that most of the PageRank trends done in the network was influenced by outflowing nodes. The inward nodes, on the other hand, as shown in Figure 11, show that they less influenced the PageRank trends because they were only able to explain 32% and 29% of the PageRank strength and PageRank, respectively. In other words, the outflowing links are more critical to the performance of the road network compared to the inflowing links.

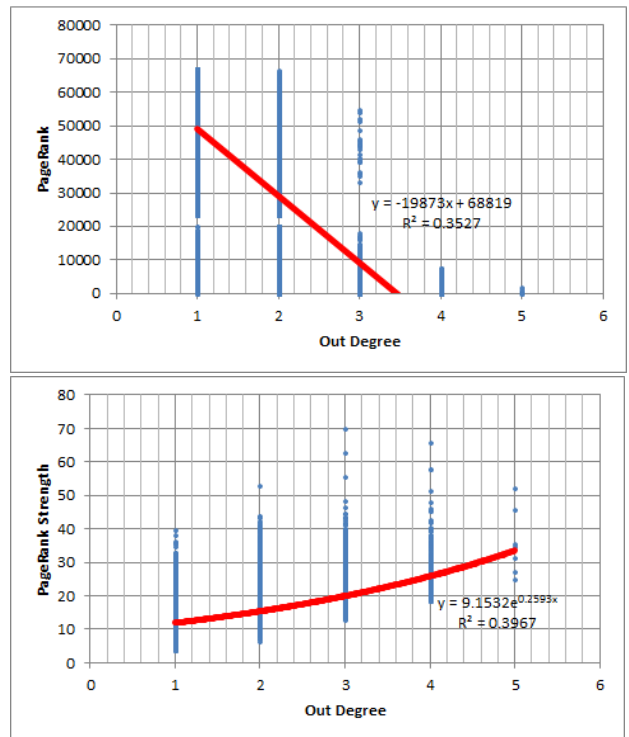

**Fig. 10** Out-degree Plot Against PageRank and PageRank Strength

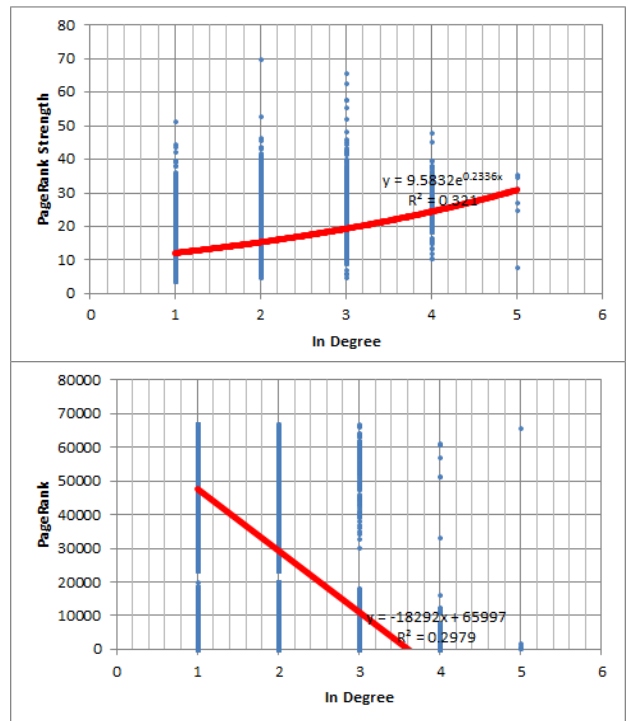

**Fig. 11** In-degree Plot Against PageRank and PageRank strength

The relationship between the PageRank and PageRank strength of the total degrees had R-values of 29% and 36%, respectively, as shown in Figure 12. This means that the total nodes influenced the PageRank and PageRank strength in accordance to the in-degree and out-degree. The three analyses were consistent to one another, meaning that as the PageRank strength increased, the number of links, whether in-degree or out-degree, increased as well. This implies that the reason why a road is considered critical because of the number of links it is connected to. A lower PageRank meant that less links were connected, specifically the outflowing links. This goes for both comparison with PageRank and PageRank strength, which explains Brin and Page's assumption that more important nodes are more likely to receive more links [3].

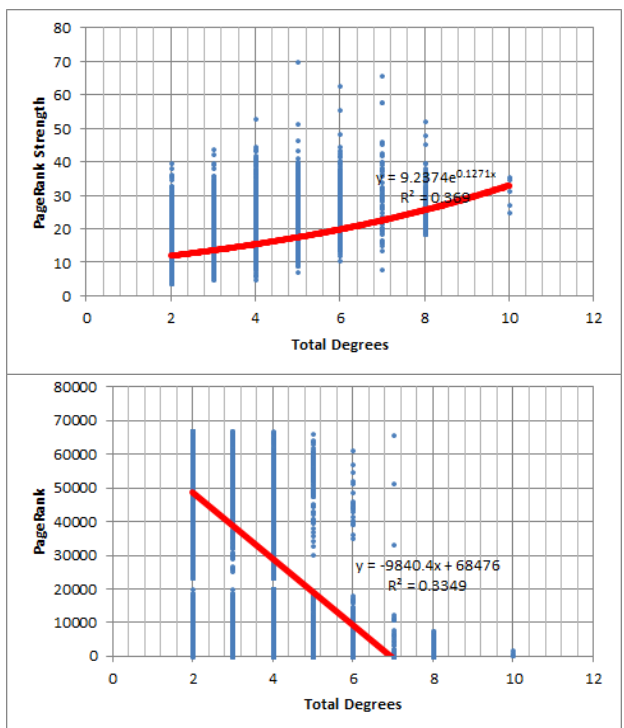

**Fig. 12** All-degree Plot Against PageRank and PageRank Strength

The lower ranking hubs had lesser strength compared to the higher-ranking hubs because the out-degrees for lower ranking hubs were only limited to 1. On the other hand, higher-ranked hubs were significantly stronger, as we have shown in the study, because they had multiple out-degrees and in-degrees. This shows that PageRank was effective in classifying important hubs in the network. Figure 14 shows that one segment of the network in which least-ranked nodes only had at least 1 out-degree or in-degree, while higher-ranked nodes had 3 or 4 links connected to them, with more outflowing links than inflowing links.

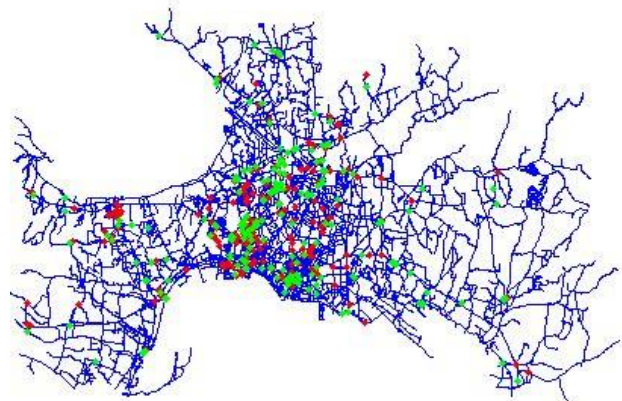

**Fig. 13** Node Ranks of 250 Highest and Lowest Ranked Nodes in Metro Manila Network

Based on Figure 13, the red markers, which signify the 250 highest-ranking hubs, were situated almost the same locations similar to the 250 lowest-ranking nodes. Upon looking at the detail, however, this is not the case.

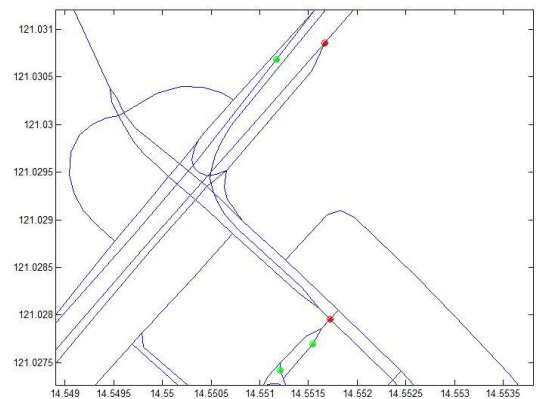

**Fig. 14** Sample Higher-Ranked Nodes (Red) with Lesser-Ranked Nodes (Green)

In detail, we found out that throughout Metro Manila, the highest and lowest ranked hubs were located at different locations, with the higher ranked hubs situated in intersections and T-junctions, while the lowest ranked nodes situated in connecting nodes between two links in one-way routes or corner roads. Figure 14 shows an example of the detail locations of nodes with the highest and lowest ranked nodes being emphasized in red and in green dots, respectively.

## 5. Conclusion

We conducted a road network analysis by identifying the most important hubs in a road network using PageRank based on topology. We propose this model to identify which nodes in the road network would be prone to possible terrorist attacks. Based on topology alone, we were able to identify the different hubs in the case study of the Metro Manila road network. We found out that the ranking trend is a sigmoid-like function, where the highest-ranking hubs were highly important compared to the lower-ranking hubs. We found out that the capital city of Manila has the most number of highest-ranking hubs. Furthermore, most high-ranking hubs were located in public places, specifically in highways, expressways, business districts, schools, train stations, and even in the port and airport.

We used topology-based PageRank and local knowledge of the places in Metro Manila. We found out that intersections and T-junction roads were the most critical hubs, while the roads with only one outflow was considered the least important hubs.

Possible future work for this research may include flow and capacity analysis with PageRank to assess vehicle behavior. We would also like to analyze road network topology using PageRank through road reconstruction. We are also looking to apply our model to different road networks, provided that we have local knowledge of the roads that we are studying. Our model is applicable to any kind of road network, not just in the Manila network. But the gathered data will be from the OSM.

## Acknowledgement

We would like to acknowledge Leandro Isla for providing the data from OSM used for our research. We would also like to give acknowledgement to DOST-ERDT for sponsoring the primary author's scholarship program, as well as CHED-PHERNET for the research fund.